\documentclass[11pt,showpacs,preprintnumbers,superscriptaddress,amsmath,amssymb,nofootinbib]{revtex4}
\usepackage{graphicx}
\usepackage{dcolumn}
\usepackage{bm}
\usepackage{amssymb}
\usepackage{amsmath}
\usepackage{epsfig}    
\usepackage{color}
\usepackage{slashed}
\usepackage{hhline}

\def\be{\begin{equation}}
\def\ee{\end{equation}}
\newcommand{\bea}{\begin{eqnarray}}
\newcommand{\eea}{\end{eqnarray}}
\newcommand{\nn}{\nonumber}

\begin{document}


\title{Texture zeros realization in a three-loop radiative neutrino mass model from modular $A_4$ symmetry}

\preprint{KYUSHU-HET-268}

\author{Takaaki Nomura}
\email{nomura@scu.edu.cn}
\affiliation{College of Physics, Sichuan University, Chengdu 610065, China}

\author{Hiroshi Okada}
\email{okada.hiroshi@phys.kyushu-u.ac.jp}
\affiliation{Department of Physics, Kyushu University, 744 Motooka, Nishi-ku, Fukuoka, 819-0395, Japan}

\author{Hajime Otsuka}
\email{otsuka.hajime@phys.kyushu-u.ac.jp}
\affiliation{Department of Physics, Kyushu University, 744 Motooka, Nishi-ku, Fukuoka, 819-0395, Japan}

\date{\today}

\begin{abstract}
We derive verifiable two-zero textures in the framework of a three-loop induced neutrino mass model applying a modular $A_4$ symmetry for the lepton sector.
Interestingly, the neutrino mass structure is determined by assignments of the right-handed charged-lepton particles only.
We show all the possible texture patterns and demonstrate their analytical and numerical analyses.
Finally, we mention a possibility of detection for doubly-charged boson via colliders that decays into specific modes due to the flavor symmetry. 

 %
 \end{abstract}
\maketitle
\newpage

\section{Introduction}
It is important to understand a mechanism of large mixings, CP phases, and tiny neutrino masses in the lepton sector since clarifying the origin sometimes leads us to obtain the main source of Baryon Asymmetry of Universe via leptogenesis~\cite{Fukugita:1986hr} which would not be explained by a framework of the standard model (SM) even after the discovery of the SM Higgs boson. Moreover, exploring the lepton mixings  would  have to confront flavor physics such as lepton flavor violations (LFVs) and muon $g-2$
once we fix a mechanism of generating nonzero neutrino masses. 

Radiative seesaw mechanisms would be one of the verifiable scenarios by current experiments because their energy scale is typically TeV and the SM leptons have interactions with exotic particles inside the loops.
These kinds of new fields potentially provide us with rich phenomenologies such as a dark matter (DM) candidate and induce lepton flavor physics depending on models.
In order to understand and predict the lepton mixings, we have useful tools; flavor dependent groups.
Abelian symmetries typically serve us zero textures in neutrino mass matrix that give us several predictions. For example, a gauged $U(1)_{L_\mu-L_\tau}$ symmetry is well known as a promising candidate to get the zero textures and explains lepton flavor physics such as muon $g-2$~\cite{Ma:2001md, Baek:2001kca, Nagao:2022osm}. On the other hand, non-Abelian symmetries typically supply predictable but non-trivial mass matrices~\cite{Altarelli:2010gt, Ishimori:2010au, Ishimori:2012zz} where they give relations among components in the mass matrices. 

In this paper, we apply a modular $A_4$ group in lepton sector~\cite{Feruglio:2017spp, Criado:2018thu, Kobayashi:2018scp, Okada:2018yrn, Nomura:2019jxj, Okada:2019uoy, deAnda:2018ecu, Novichkov:2018yse, Nomura:2019yft, Okada:2019mjf,Ding:2019zxk, Nomura:2019lnr,Kobayashi:2019xvz,Asaka:2019vev,Zhang:2019ngf, Gui-JunDing:2019wap,Kobayashi:2019gtp,Nomura:2019xsb, Wang:2019xbo,Okada:2020dmb,Okada:2020rjb, Behera:2020lpd, Behera:2020sfe, Nomura:2020opk, Nomura:2020cog, Asaka:2020tmo, Okada:2020ukr, Nagao:2020snm, Okada:2020brs,Kang:2022psa} and generate some two zero textures. It implies that we do not use triplet representation in $A_4$ but three singlets only.  Thanks to additional degrees of freedom that is modular weight, we can obtain predictable textures in a three loop model~\cite{Gustafsson:2012vj}.\footnote{See for a different approach to realize texture zeros in quark and lepton mass matrices using the modular symmetry \cite{Lu:2019vgm,Kikuchi:2022svo,Ding:2022aoe,Devi:2023vpe}.} Interestingly, the forms of texture are determined via assignments of right-handed charged-lepton particles and nonzero modular weights are imposed in new fields only. After showing analytical and numerical predictions, we briefly mention a possibility of detectability at colliders because flavor specific signals are expected in exotic fields.

This paper is organized as follows.
In Sec. \ref{sec:II}, we review two zero textures that can satisfy the current neutrino oscillation data.
In Sec. \ref{sec:III}, we show the field contents and their assignments and derive two zero textures in the neutrino mass matrix. Then, we demonstrate analytical formulae and discuss whether two zero textures satisfy the current neutrino oscillation data. 
In Sec. \ref{sec:IV}, we give numerical analysis and show some predictions for each case. 
And then, we briefly mention  a possibility of testing models at collider experiments.
 Finally, we summarize and conclude in Sec. \ref{sec:V}.

\section{Neutrino two zero textures}
\label{sec:II}
In this section, we review the two zero texture of neutrino sector based on a diagonal charged-lepton mass matrix, supposing the active neutrinos are Majorana fermions.~\footnote{Any three zero textures do not satisfy the current neutrino oscillation data that is shown in e.g., ref.~\cite{Fritzsch:2011qv}. }
In this case, the PMNS matrix denoted by $U_{PMNS}$~\cite{Maki:1962mu, Pontecorvo:1967fh} originates from the neutrino sector only.
It is known that the two zero textures are the minimum ones to satisfy the current neutrino oscillation data.
The neutrino mass matrix denoted by $M_\nu$ is symmetric due to the Majorana property, and one straightforwardly finds that there exist fifteen patterns in total. However, only seven patterns are allowed to reproduce the current data as follows~\cite{Fritzsch:2011qv}:
\begin{align}
&\left[\begin{array}{ccc}
0 & 0 & \times \\
0 & \times & \times \\
\times & \times & \times \\
\end{array}\right],\quad 
\left[\begin{array}{ccc}
0 & \times & 0 \\
\times & \times & \times \\
0 & \times & \times \\
\end{array}\right],\quad
\left[\begin{array}{ccc}
\times & \times & 0 \\
\times & 0 & \times \\
0 & \times & \times \\
\end{array}\right],\quad
\left[\begin{array}{ccc}
\times & 0 & \times \\
0 & \times & \times \\
\times & \times & 0 \\
\end{array}\right],\\
&\left[\begin{array}{ccc}
\times & 0 & \times \\
0 & 0 & \times \\
\times & \times & \times \\
\end{array}\right],\quad
\left[\begin{array}{ccc}
\times & \times & 0 \\
\times & \times & \times \\
0 & \times & 0 \\
\end{array}\right],\quad
\left[\begin{array}{ccc}
\times & \times & \times \\
\times & 0 & \times \\
\times & \times & 0 \\
\end{array}\right],
\end{align}
where the symbol of $"\times"$ indicates a nonzero component.
These textures can be diagonalized by $U_{PMNS}$ as follows:
\begin{align}
M_\nu=U_{PMNS}  
\left[\begin{array}{ccc}
m_1 & 0 & 0 \\
0 & m_2 & 0 \\
0 & 0 & m_3 \\
\end{array}\right]
U_{PMNS}^T,
\end{align}
where $m_{1,2,3}$ is the neutrino mass eigenvalues, and $U_{PMNS}$ can be decomposed by $V\cdot P$, where $V$ is a three by three unitary matrix consisting of three mixing angles $\theta_{12,23,13}$ and one Dirac CP-phase $\delta$.
$P$ is  a phase matrix given by $\rm{diag.}(e^{i\rho},e^{i\sigma},1)$, where $\rho,\ \sigma$ are Majorana phases.
~\footnote{After rotating the $M_\nu$ by $V$, the mass eigenvalues are generally complex. But one phase can be removed by rotating the whole neutrino eigenstates. In this case, we removed the phase of third mass eigenvalue following the ref.~\cite{Fritzsch:1997st}.}
Similar to the quark sector, $V$ is explicitly written by~\footnote{This form has a different location of Dirac CP phase from the standard parametrization in Particle Data Group~\cite{ParticleDataGroup:2014cgo}, but it is not crucial for observables due to phase redefinitions of quarks~\cite{Fritzsch:1997st}. In fact, one finds this form by redefining $\mu\to \mu e^{i\delta}$, $\tau\to \tau e^{i\delta}$ and $\nu_\tau\to\nu_\tau  e^{i\delta}$ from the standard form. }
\begin{align}
V= 
\left[\begin{array}{ccc}
c_{12}c_{13} & s_{12}c_{13} & s_{13} \\
-c_{12} s_{23} s_{13}-s_{12}c_{23}e^{-i\delta} & -s_{12} s_{23} s_{13}+c_{12}c_{23}e^{-i\delta} & s_{23} c_{13} \\
-c_{12} c_{23} s_{13}+s_{12}s_{23}e^{-i\delta} & -s_{12} c_{23} s_{13}-c_{12}s_{23}e^{-i\delta} & c_{23} c_{13} \\
\end{array}\right],
\end{align}
where $s_{ij},\ c_{ij}$ are respectively  short-hand notations for $\sin\theta_{ij},\ \cos\theta_{ij}$ ($i,j=12,23,13$).
In our convenient notation later, we rewrite the mass eigenvalues with complex values as follows:
 \begin{align}
M_\nu=V  
\left[\begin{array}{ccc}
e^{i\rho} & 0 & 0 \\
0 & e^{i\sigma} & 0 \\
0 & 0 & 1 \\
\end{array}\right]
\left[\begin{array}{ccc}
m_1 & 0 & 0 \\
0 & m_2 & 0 \\
0 & 0 & m_3 \\
\end{array}\right]
\left[\begin{array}{ccc}
e^{i\rho} & 0 & 0 \\
0 & e^{i\sigma} & 0 \\
0 & 0 & 1 \\
\end{array}\right]
V^T
\equiv
V 
\left[\begin{array}{ccc}
\lambda_1 & 0 & 0 \\
0 & \lambda_2 & 0 \\
0 & 0 & \lambda_3 \\
\end{array}\right]
V^T,
\end{align}
 where $(\lambda_1,\lambda_2,\lambda_3)\equiv (m_1 e^{2i\rho}, m_2 e^{2i\sigma}, m_3 )$.
 Then, applying that the neutrino mass matrix has two zero textures denoted by $(M_\nu)_{ab}=(M_\nu)_{cd}=0$ ($ab\neq cd$),
 one straightforwardly derives the following relations:
 \begin{align}
 \Lambda_{13}&\equiv \frac{\lambda_1}{\lambda_3}= 
 \frac{V_{a3} V_{b3} V_{c2} V_{d2} - V_{a2} V_{b2} V_{c3} V_{d3}}{V_{a2} V_{b2} V_{c1} V_{d1} - V_{a1} V_{b1} V_{c2} V_{d2}},\nn\\
 \Lambda_{23}&\equiv \frac{\lambda_2}{\lambda_3}= 
 \frac{V_{a1} V_{b1} V_{c3} V_{d3} - V_{a3} V_{b3} V_{c1} V_{d1}}{V_{a2} V_{b2} V_{c1} V_{d1} - V_{a1} V_{b1} V_{c2} V_{d2}}.
 \label{eq:relori}
 \end{align}
Furthermore, one finds the following mass ratios and Majorana phases: 
 \begin{align}
\xi \equiv \frac{m_1}{m_3}&=\left| \Lambda_{13} \right|,\
\zeta \equiv \frac{m_2}{m_3} =\left| \Lambda_{23}\right|,\
\rho =
\frac12{\rm arg}\left[ \Lambda_{13} \right],\
\sigma =
\frac12{\rm arg}\left[ \Lambda_{23} \right].
 \end{align}
 Two mass square differences $\Delta m^2_{\mathrm{sol}}\equiv m_2^2-m_1^2$ and $\Delta m^2_{\mathrm{atm}}\equiv m_3^2-(m_2^2+m_1^2)/2$
are rewritten in terms of $\xi, \zeta, m_3$ as follows:
\begin{align}
& \Delta m^2_{\mathrm{sol}} = m_3^2(\zeta^2-\xi^2),\
\Delta m^2_{\mathrm{atm}} = m_3^2(1-(\zeta^2+\xi^2)/2).
 \end{align}
In addition, the effective mass for the neutrinoless double beta decay is given by
\begin{align}
\langle m_{ee}\rangle=\left|\sum_{i=1}^3 m_i (U_{PMNS})_{ei}^2\right|,
\end{align}
where its predicted value is constrained by the current KamLAND-Zen data and could be measured in future~\cite{KamLAND-Zen:2016pfg,KamLAND-Zen:2022tow}.
The upper bound is found as $\langle m_{ee}\rangle<(61-165)$ meV at 90 \% confidence level where the range of the bound comes from the use of different method estimating nuclear matrix elements. 
Sum of neutrino masses is constrained by the minimal cosmological model
$\Lambda$CDM $+\sum m_i$ that provides the upper bound on $\sum m_i\le$ 120 meV~\cite{Vagnozzi:2017ovm, Planck:2018vyg}, although it becomes weaker if the data are analyzed in the context of extended cosmological models~\cite{ParticleDataGroup:2014cgo}.
These two observables $\langle m_{ee}\rangle$ and $\sum m_i$ are also taken into account in the numerical analysis.

 Inserting the experimental mixing angles via Nufit 5.2~\cite{Esteban:2020cvm}, $\Lambda_{13}$ and  $\Lambda_{23}$ are uniquely fixed. Then, one can check two mass square differences are within the experimental ranges at 3 $\sigma$ and predict Majorana phases.
 In the next section, we will show an intriguing example of model building based on a three-loop neutrino mass model~\cite{Gustafsson:2012vj} in a modular $A_4$ flavor symmetry.

\section{Model setup}
\label{sec:III}

\begin{table}[t!]
\begin{tabular}{|c||c|c|c|c|c|c|c|c|c|c|c|}\hline\hline  
& ~$\hat L$~ & ~$\hat {\overline{\ell}}$~ & ~$\hat H_{u}$~ & ~$\hat H_{d}$~ & ~$\hat \eta_{u}$~ & ~$\hat \eta_{d}$~ & ~$\hat k^{++}$~ & ~$\hat k^{--}$~ 
& ~$\hat s^+$~ & ~$\hat s^-$~& ~$\hat \chi$~  \\\hline\hline 
$SU(2)_L$   & $\bm{2}$  & $\bm{1}$  & $\bm{2}$ & $\bm{2}$ & $\bm{2}$  & $\bm{2}$ & $\bm{1}$ & $\bm{1}$  & $\bm{1}$  & $\bm{1}$ & $\bm{1}$     \\\hline 
$U(1)_Y$    & $-\frac12$  & $+1$ & $+\frac12$ & $-\frac12$ & $+\frac12$  & $-\frac12$ & $+2$ & $-2$ & $+1$ & $-1$ & $0$ \\\hline
$A_4$   & $\{\bm{1}\}$  & $\{ \overline{\bm{1}}\}$  & $\bm{1}$ & $\bm{1}$ & $\bm{1}$  & $\bm{1}$ & $\bm{1}$ & $\bm{1}$  & $\bm{1}$& $\bm{1}$& $\bm{1}$     \\\hline 
$-k_I$    & $0$  & $0$ & $0$ & $0$ & $-2$ & $-2$ & $-4$& $-4$ & $-2$ & $-2$ & $-2$   \\\hline
\end{tabular}
\caption{Charge assignments of the SM lepton and new superfields
under $SU(2)_L\times U(1)_Y \times A_4$ where $-k_I$ is the number of modular weight.}\label{tab:1}
\end{table}

In this section, we introduce a three-loop neutrino mass model based on a modular $A_4$ flavor symmetry.
To forbid infinite terms, we work on a supersymmetric scenario 
but we only use the SM fields and new bosons in our model. The superfields are denoted by $\hat \psi$
and we assign nonzero modular weights to new superfields only that we introduce. 
In our new superfield sector, we introduce two isospin doublet inert ones $\hat \eta_{u,d}$, two isospin singlet doubly-charged ones $\hat k^{\pm\pm}$, two isospin singlet singly-charged ones $\hat s^{\pm}$, and an isospin singlet neutral one $\hat\chi$ in addition to the MSSM Higgs superfields $\hat H_{u,d}$.
All these new superfields are assigned by trivial singlet under $A_4$ but nonzero modular weights are assigned for new fields as shown in Table~\ref{tab:1}.
We then have associated scalar bosons from these superfields that play a role in generating neutrino masses.
As for the SM matter superfield sector, we assign three different types of singlets $1,1',1''$ under the $A_4$ to the left-handed leptons $\hat L\equiv[\hat\nu,\hat\ell]^T$ and right-handed $\hat{\overline {\ell}}$ where we fix the $A_4$ representations such that charged-lepton mass matrix is always diagonal. Thus there are six patterns of assigning their representations to the charged leptons as we will discuss later. 

The renormalizable superpotential which are invariant under the modular $A_4$ is found as
\begin{align}
W_\ell = & 
y_\ell \hat{\overline{\ell}} \hat L_\ell \hat H_d + [\hat{\bar \ell} g_{\ell\ell'} \hat{\overline {\ell'}}^T] \hat k^{--}
+ \mu_{ssk} \hat s^+ \hat s^+ \hat k^{--} + \mu'_{ssk} \hat s^- \hat s^- \hat k^{++}
+ \mu_{h\eta s} \hat H_u \hat \eta_u \hat s^{-}  
+ \mu'_{h\eta s} \hat H_d \hat \eta_d \hat s^{+} \nonumber \\
&
+ \mu_{h\eta \chi} \hat H_u \hat \eta_d \hat \chi  
+ \mu'_{h\eta \chi} \hat H_d \hat \eta_u \hat \chi  
+ \mu_{\chi} \hat \chi \hat \chi
+ \mu_H \hat H_u \hat H_d
+ \mu_\eta \hat \eta_u \hat \eta_d
+ \mu_s \hat s^+ \hat s^-,
\label{eq:SP}
\end{align}
where the R-symmetry is implicitly imposed.
Note that $A_4$ singlets modular forms are implicitly included in the coupling constants to make terms invariant under modular symmetry. Since singlet modular forms are just a complex value, we can absorb them in coupling constants; e.g. $g_{\ell \ell'} \equiv Y^{(4)}_{1,1'} \hat{g}_{\ell \ell'}$ where $Y^{(4)}_{1,1'}$ is the $A_4$ singlet modular forms with weight 4 and $\hat{g}_{\ell \ell'}$ is an original coupling constant. 
Moreover, we need the following current interaction
\begin{align}
\frac{g}{\sqrt{2}}\bar\ell \gamma^\mu 
P_L \nu W^-_\mu + \mathrm{h.c.},
\end{align}
where $\nu$ is flavor eigenstate and $g$ denotes the $SU(2)$ gauge coupling. 
Such an interaction can be derived from the K\"ahler potential:
\begin{align}
   K= \hat{L}^\ast e^{V} \hat{L}
\end{align}
where $V$ denotes the vector multiplet of the Standard Model.

To generate the neutrino masses, we need scalar mixings that can be obtained by the following soft supersymmetry (SUSY) breaking terms: 
\begin{align}
&A_{ssk}  s^+  s^+  k^{--} + A'_{ssk}  s^-  s^-  k^{++}
+ A_{h\eta s}  H_u  \eta_u  s^{-}  
+ A'_{h\eta s}  H_d  \eta_d  s^{+}  
+ A_{h\eta \chi}  H_u  \eta_d  \chi\nn\\  
&+ A'_{h\eta \chi}  H_d  \eta_u  \chi  
+ B_H^2  H_u  H_d
+ B_\eta^2 \eta_u  \eta_d
+ B_s^2  s^+  s^-
+ B_\chi^2 \chi \chi +{\rm h.c.}.
\end{align}
Notice that the fields without hat in the equation are scalar components associated with each superfield. In addition, $A_4$ singlet modular forms are implicitly included in coupling constants if they are necessary to make term invariant under modular symmetry as in the superpotential.\footnote{See for a realization of modular-invariant soft SUSY breaking terms \cite{Kobayashi:2021uam,Kikuchi:2022pkd}.}

\subsection{Neutrino mass matrix}
\begin{figure}[tb]
\begin{center}
\includegraphics[width=100.0mm]{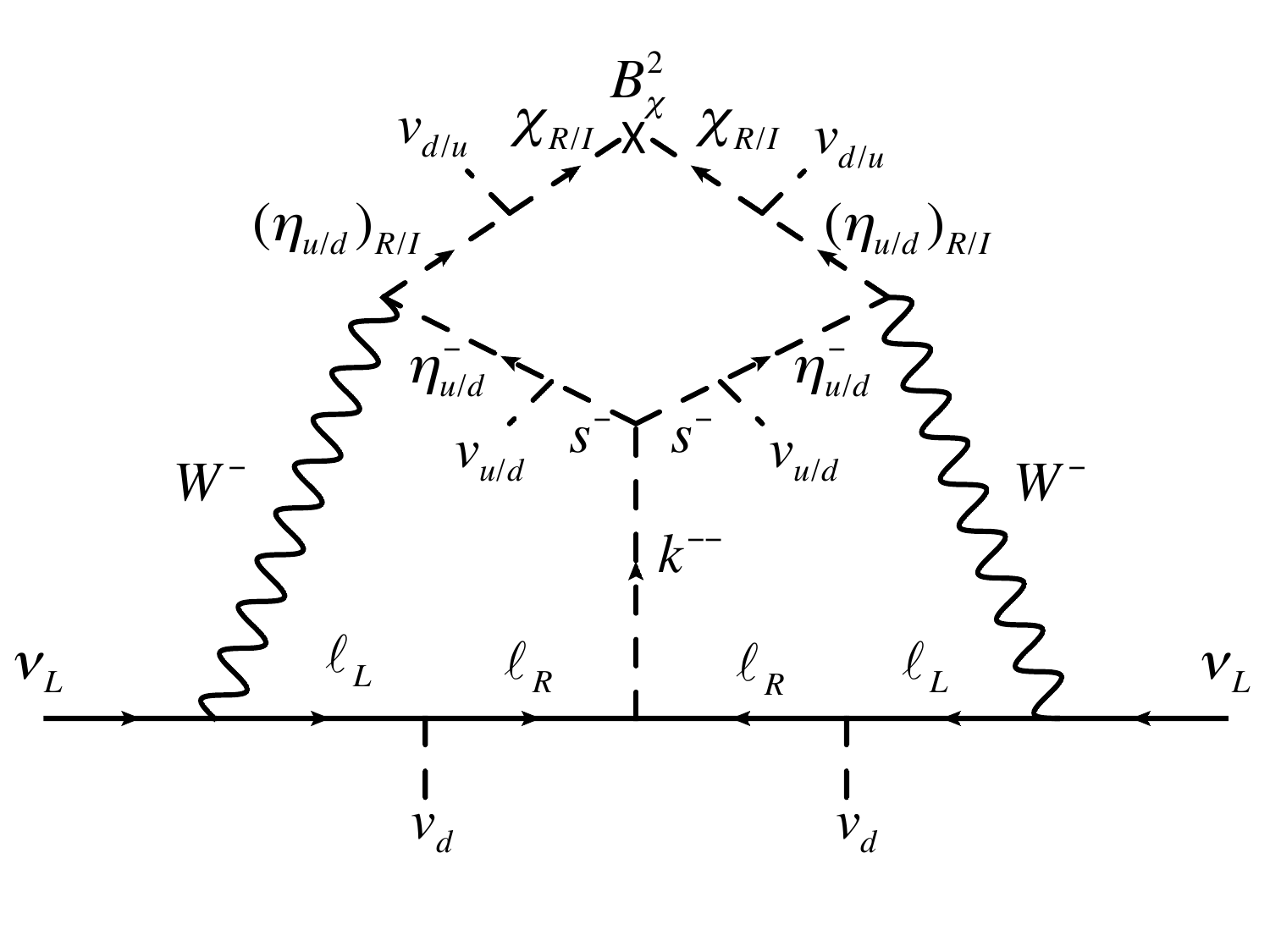} 
\caption{One loop diagram generating active neutrino mass matrix. }
  \label{fig-neut}
\end{center}\end{figure}

In this subsection, the neutrino mass matrix at three-loop level is given via Fig.~\ref{fig-neut}
\begin{align}
(m_\nu)_{ab}=\frac{{\cal M}_{0}}{(16\pi^2)^3} y_\ell g_{\ell\ell'} y_{\ell'} ,
\end{align}
where ${\cal M}_{0}$ is a loop function with one mass dimension that is written in terms of mass eigenvalues of bosons inside the loop. The detailed formula of loop function is given in ref.~\cite{Gustafsson:2012vj}. Notice here that the loop function does not depend on the structure of neutrino mass matrix since all the charged-lepton masses are negligible compared to the other running masses inside the loop, therefore, {\it the neutrino mass structure depends on that of $g_{\ell\ell'}$ only.} 
Here, we concentrate on the neutrino mass texture below where the texture is determined by the assignment for the right-handed charged-leptons under the modular $A_4$.

\subsection{Possible matrices of $g$ of assignments for the right-handed charged-leptons}
Depending on assignments of the modular $A_4$ for the right-handed charged-leptons $(\bar e, \bar\mu,\bar\tau)$, we find three possible two-zero textures of the neutrino mass matrix which are summarized as follows.
\begin{align}
&(1) \  (1,1'',1'),\  (1',1'',1): \left(\begin{array}{ccc} \times & \times & 0 \\ \times & 0 & \times \\ 0 & \times & \times \end{array} \right), \quad
(2) \  (1,1',1''),\  (1',1,1'') : \left( \begin{array}{ccc} \times & 0 & \times \\ 0 & \times & \times \\ \times & \times & 0 \end{array} \right), \nonumber \\
&(3) \  (1'',1,1'),\  (1'',1',1) : \left( \begin{array}{ccc} 0 & \times & \times \\ \times & \times & 0 \\ \times & 0 & \times \end{array} \right).
\end{align}
These two-zero textures can be realized since weight 4 modular form cannot be $1''$ under $A_4$; coupling $g_{\ell \ell'}$ should include weight 4 modular form as we mentioned below Eq.~\eqref{eq:SP}. 
Thus a matrix element vanishes if $1''$ modular form is required for invariance under modular $A_4$ symmetry. 
The last texture, $(3)$, does not satisfy the current neutrino oscillation data as shown below, and we demonstrate numerical analyses for the textures of $(1)$ and $(2)$ in the next section. 
Notice here that the matrix (1) is related to a permutation matrix $P_{23}$:
\begin{align}
M_1 = P_{23} M_2 P^T_{23},\quad
P_{23} \equiv \left( \begin{array}{ccc} 1 & 0 & 0 \\ 0 & 0 & 1 \\ 0 & 1 & 0 \end{array} \right), 
\end{align}
where $M_{1}$ and $M_2$ are respectively the matrices of $(1)$ and $(2)$.
It suggests that PMNS matrix can be redefined by $P_{23}\cdot  V\cdot P$, therefore, we straightforwardly find the following relations:
\begin{align}
\theta_{12}^{(2)}=\theta_{12}^{(1)},\quad
\theta_{13}^{(2)}=\theta_{13}^{(1)},\quad
\theta_{23}^{(2)}=\frac{\pi}{2}-\theta_{23}^{(1)},\quad
\delta^{(2)}= \delta^{(1)}-\pi. \label{eq:rel12}
\end{align}
where each the upper index represents the case of $(1)$ or $(2)$.
Once one finds relations for either of the cases, one obtains the other relations by substituting Eq.~(\ref{eq:rel12}).

We will adopt the neutrino experimental data at 3$\sigma$ interval in Nufit 5.2~\cite{Esteban:2020cvm} as follows:
\begin{align}
&{\rm NH}: \Delta m^2_{\rm atm}=[2.428, 2.597]\times 10^{-3}\ {\rm eV}^2,\
\Delta m^2_{\rm sol}=[6.82, 8.03]\times 10^{-5}\ {\rm eV}^2,\label{exp:nh}\\
&\sin^2\theta_{13}=[0.02029, 0.02391],\ 
\sin^2\theta_{23}=[0.406, 0.620],\ 
\sin^2\theta_{12}=[0.270, 0.341],\
\delta=[108^\circ, 404^\circ],\nn\\
&\theta_{13}/\circ=[8.19,8.89],\quad \theta_{23}/\circ=[39.6, 51.9],\quad \theta_{12}/\circ=[31.31,35.74],\nn\\
&{\rm IH}: \Delta m^2_{\rm atm}=[2.408, 2.581]\times 10^{-3}\ {\rm eV}^2,\
\Delta m^2_{\rm sol}=[6.82, 8.03]\times 10^{-5}\ {\rm eV}^2,\label{exp:ih} \\
&\sin^2\theta_{13}=[0.02047, 0.02396],\ 
\sin^2\theta_{23}=[0.412, 0.623],\ 
\sin^2\theta_{12}=[0.270, 0.341],\
\delta=[192^\circ, 360^\circ],\nn\\
& \theta_{13}/\circ=[8.23,8.90],\quad \theta_{23}/\circ=[39.9, 52.1],\quad \theta_{12}/\circ=[31.31,35.74].
\nn
\end{align}

\subsubsection{Case (1) }
At first, we explicitly write down the exact relations of Eqs.~(\ref{eq:relori}) in the case of (1) putting $(M_\nu)_{13}=(M_\nu)_{22}=0$ as follows:
\begin{align}
\frac{\lambda_1}{\lambda_3} &= \frac{s_{12}c_{12}s_{23}(2c^2_{23}s^2_{12}-s^2_{23}c^2_{13}) -c_{23}s_{13}(c^2_{12}c^2_{23}e^{-i\delta}+c^2_{12}c^2_{23}e^{-i\delta})}{s_{12}c_{12}s_{23}c^2_{23} +(s^2_{12}-c^2_{12})c^3_{23}s_{13}e^{i\delta} +s_{12}c_{12}s_{23}s^2_{13}(1+c^2_{23})e^{2i\delta}}
e^{2i\delta},\\
\frac{\lambda_2}{\lambda_3} &= \frac{s_{12}c_{12}s_{23}(2c^2_{23}s^2_{12}-s^2_{23}c^2_{13}) +c_{23}s_{13}(c^2_{12}s^2_{23}e^{i\delta}+s^2_{12}c^2_{23}e^{-i\delta})}{s_{12}c_{12}s_{23}c^2_{23} +(s^2_{12}-c^2_{12})c^3_{23}s_{13}e^{i\delta} +s_{12}c_{12}s_{23}s^2_{13}(1+c^2_{23})e^{2i\delta}}
e^{2i\delta}.
\end{align}
Considering $s_{13}$ to be small from the experimental result as already shown in Eqs.~(\ref{exp:nh},\ref{exp:ih}),
we can expand the above relations in terms of $s_{13}$. In the leading order, we find 
\begin{align}
\frac{m_1}{m_3} \approx \frac{m_2}{m_3} \approx \tan^2\theta_{23},\quad
\rho\approx\sigma \approx \delta-\frac{\pi}{2}.
\end{align}
It implies that there are no differences between $m_1/m_3$ and $m_2/m_3$, and $\rho$ and $\sigma$ at the leading order.
In order to check these differences, we consider the next leading order terms which are given by
\begin{align}
&\frac{m_1}{m_3} - \frac{m_2}{m_3} \approx \frac{4\sin\theta_{13}\cos\delta}{\sin2\theta_{12} \sin2\theta_{23}},\label{b1rel1}\\
&\rho-\sigma \approx  - \frac{2\sin\theta_{13} \sin\delta}{\sin2\theta_{12} \tan2\theta_{23}\tan^2\theta_{23}}.
\end{align}
Eq.~({\ref{b1rel1}}) might provide important information on the Dirac CP phase.
Considering the experimental results in Eqs.~({\ref{exp:nh}}) and ({\ref{exp:ih}}), 
$\sin\theta_{13}$, $\sin2\theta_{12}$, and $\sin2\theta_{23}$ are definitely positive. Therefore, $\cos\delta$ is negative; i.e. $90^\circ\le \delta\le270^\circ$, because of $m_1<m_2$. Finally, we expect that the ranges of $\delta$  in Eqs.~({\ref{exp:nh}}) and ({\ref{exp:ih}}) are restricted by
\begin{align}
{\rm NH}: 
\delta=[108^\circ, 270^\circ], \quad
{\rm IH}: 
\delta=[192^\circ, 270^\circ].\label{eq:1dirac}
\end{align}
In the next section, we numerically analyze and check the above relations.

\subsubsection{Case (2)}
Applying Eq.~(\ref{eq:rel12}) to the case (1), we straightforwardly obtained the relations as follows:
In the leading order, we find 
\begin{align}
\frac{m_1}{m_3} \approx \frac{m_2}{m_3} \approx \cot^2\theta_{23},\quad
\rho\approx\sigma \approx \delta-\frac{\pi}{2},
\end{align}
where we have used $\rho$ and $\sigma$ has a period of $\pi$.
Similar to the case (1),  there are no differences between $m_1/m_3$ and $m_2/m_3$, and $\rho$ and $\sigma$ as a leading order.
In order to check these differences, we consider the next leading order terms which are given by
\begin{align}
&\frac{m_1}{m_3} - \frac{m_2}{m_3} \approx - \frac{4\sin\theta_{13}\cos\delta}{\sin2\theta_{12} \sin2\theta_{23}},\label{b1rel2}\\
&\rho-\sigma \approx  - \frac{2\sin\theta_{13} \sin\delta}{\sin2\theta_{12} \tan2\theta_{23}\cot^2\theta_{23}}.
\end{align}
Eq.~({\ref{b1rel2}}) might provide important information on the Dirac CP phase.
Considering the experimental results in Eqs.~({\ref{exp:nh}}) and ({\ref{exp:ih}}), 
$\sin\theta_{13}$, $\sin2\theta_{12}$, and $\sin2\theta_{23}$ are definitely positive. Therefore, $\cos\delta$ is positive; i.e. $0^\circ\le \delta\le90^\circ$ and $270^\circ\le \delta\le360^\circ$, because of $m_1<m_2$. Finally, we expect that the ranges of $\delta$  in Eqs.~({\ref{exp:nh}}) and ({\ref{exp:ih}}) are restricted by
\begin{align}
{\rm NH}: 
\delta=[270^\circ, 404^\circ], \quad
{\rm IH}: 
\delta=[270^\circ, 360^\circ].
\end{align}
In the next section, we numerically analyze and check the above relations.

\subsubsection{Case (3)}
Now that the current experimental data are shown, we demonstrate the reason why the texture (3) is ruled out.
Putting $(ab)=(11)$ and  $(cd)=(23)$ in Eqs.~(6,7), $\xi$ and $\zeta$ can approximately be written by
\begin{align}
\xi&\approx -\frac{s^2_{12}}{1-2s^2_{12}},\quad
\zeta\approx \frac{1-s^2_{12}}{1-2s^2_{12}},
\end{align}
which are zeroth order of $s_{13}<<1$. 
On the other hand, a ratio of the mass squared differences is rewritten in terms of $\xi$ and $\zeta$ as follows:
\begin{align}
R=\frac{\Delta m^2_{\rm sol}}{|\Delta m^2_{\rm atm}|} =2\frac{\zeta^2 - \xi^2}{|2-(\zeta^2 + \xi^2)|}.
\end{align}
It implies that $R$ is given by
\begin{align}
R\approx \frac{2(1-2 s^2_{12})}{|1-6 s^2_{12} + 6 s^4_{12}|} .
\end{align}
Substituting $s_{12}^2=0.303$ that is the best fit value in Nufit 5.2,
we find $R\approx3$. 
While the experimental result in Eq.~(17,18) suggests that $R\approx0.03$.
 Thus, the texture (3) is totally ruled out.


\section{Numerical analysis and phenomenology}
\label{sec:IV}
In this section, we carry out numerical analysis to fit the neutrino data and find some predictions of the models.
Then some implications for phenomenology are discussed.
In addition to the experimental values in Eqs.~(17,18),  we randomly select the value of $m_3$ within the range of
\begin{align}
m_3 \in [10^{-15},10^{-9}]\ {\rm eV}.
\end{align}

\begin{figure}[tb]
\begin{center}
\includegraphics[width=70.0mm]{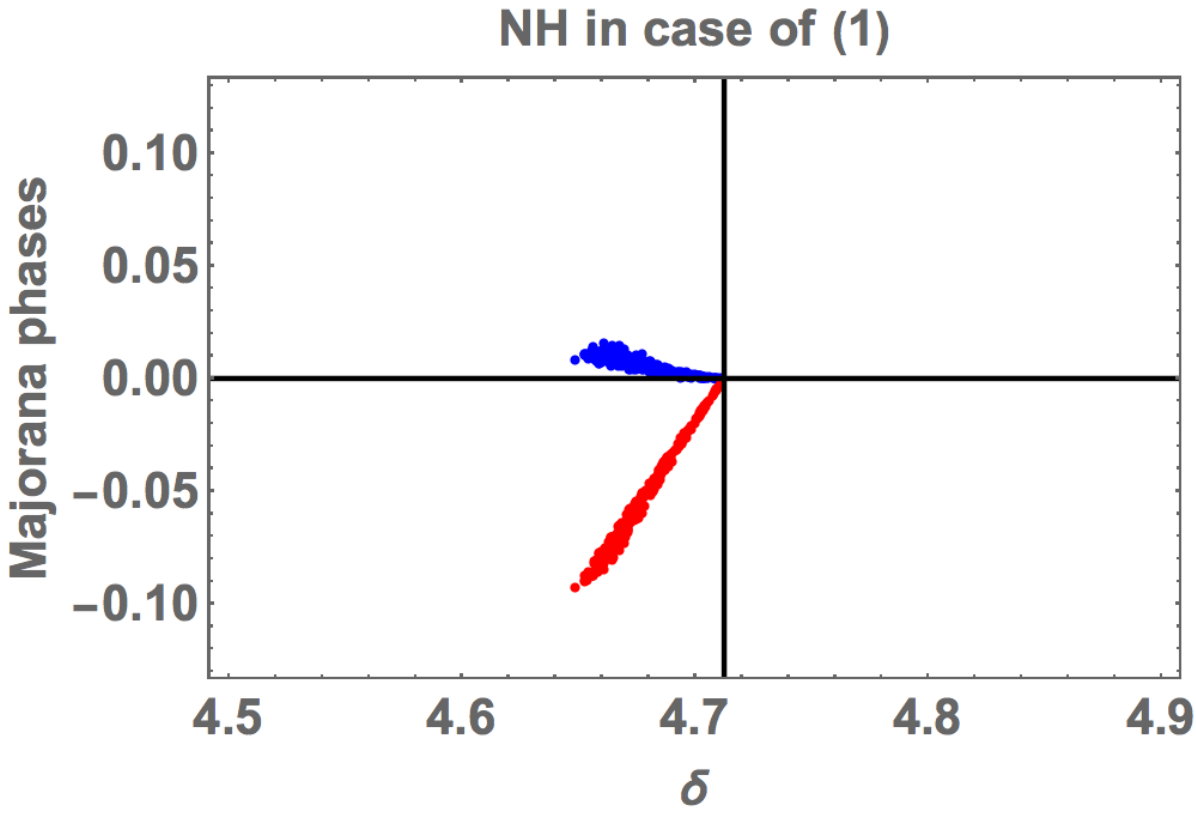} \quad
\includegraphics[width=70.0mm]{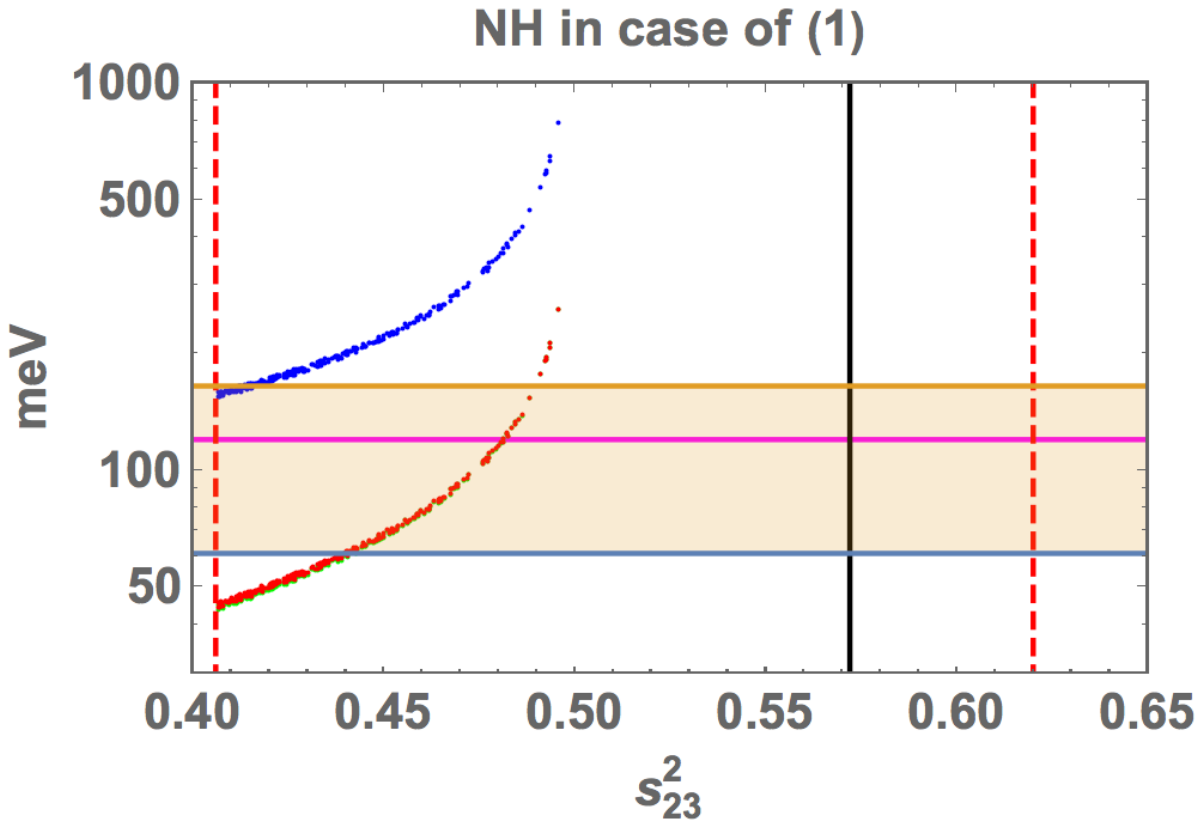}  \\
\includegraphics[width=70.0mm]{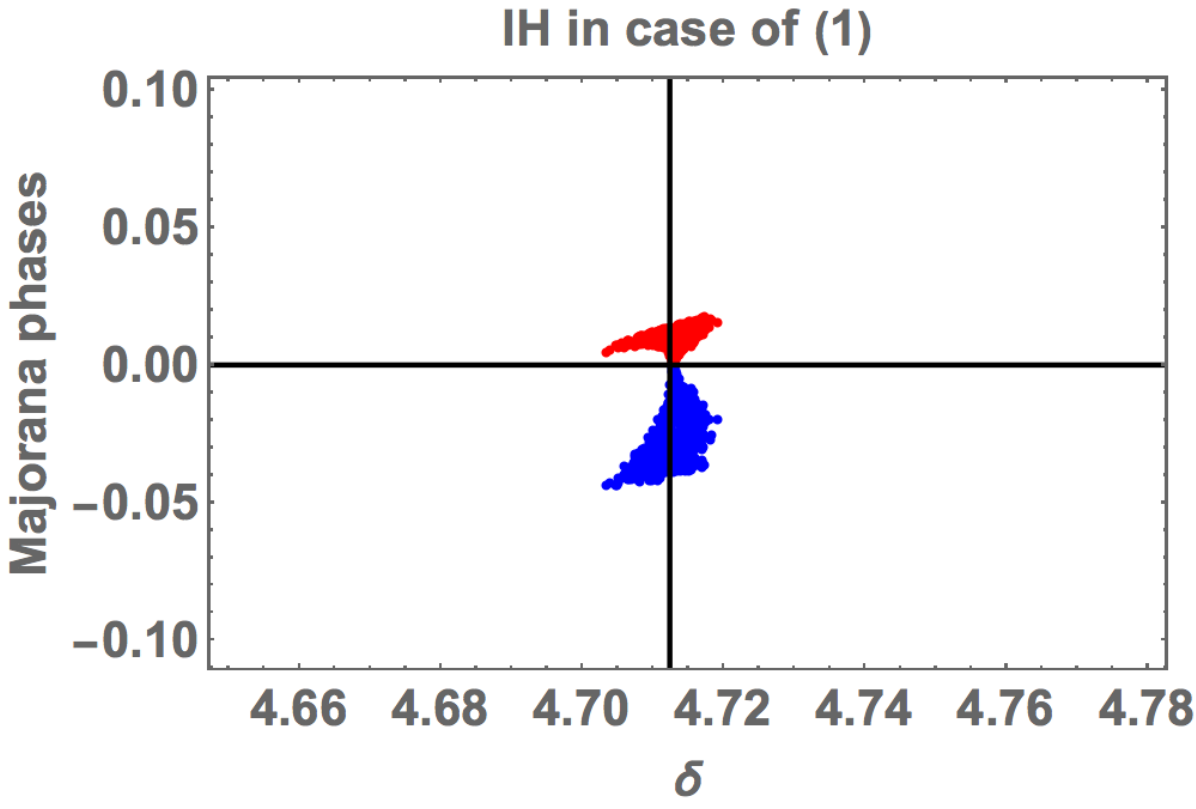} \quad
\includegraphics[width=70.0mm]{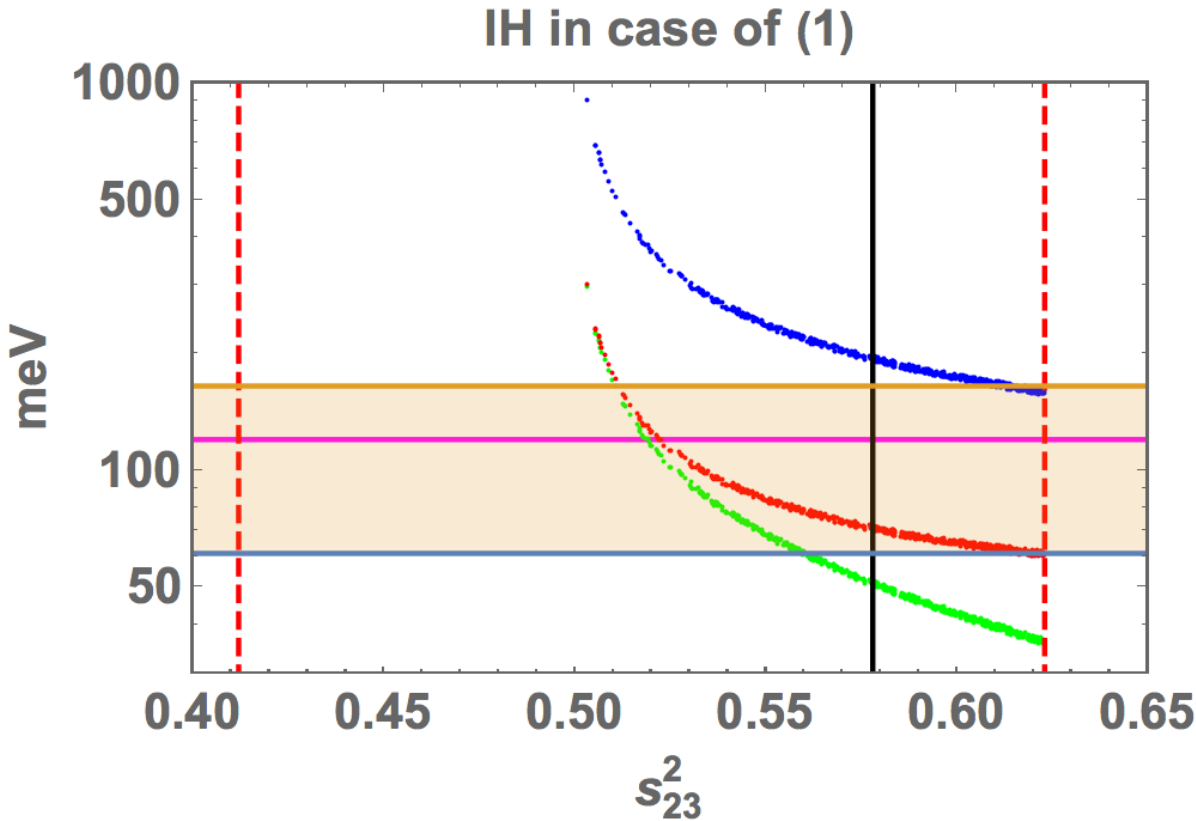} 
\caption{Numerical analyses of texture (1).
The up and down figures represent the NH and the IH cases, respectively. 
The left ones demonstrate Majorana phases $\rho$(blue) and $\sigma$(red) in terms of $\delta$,
and the right ones $\sum_im_i$(blue), $\langle m_{ee}\rangle$(red) and the lightest neutrino mass (green); $m_1$ for NH and
$m_3$ for IH, in terms of $s_{23}^2$. Each the black vertical line is $3\pi/2$ (left fig.) and BF (right fig.) and the dotted vertical red ones 3$\sigma$ interval. The horizontal pink line shows the upper bound on cosmological constant $\sum_im_i=120$ meV. The orange band corresponds to the upper bound on $\langle m_{ee}\rangle$. }
  \label{fig:B1}
\end{center}\end{figure}
%
\subsection{Texture (1)}
In Fig.~\ref{fig:B1}, we show predictions in the case of texture (1).
The up and down figures represent the NH and the IH cases, respectively. 
The left ones demonstrate Majorana phases $\rho$(blue) and $\sigma$(red) in terms of $\delta$,
and the right ones $\sum_im_i$(blue), $\langle m_{ee}\rangle$(red) and the lightest neutrino mass (green); $m_1$ for NH and
$m_3$ for IH, in terms of $s_{23}^2$. Each the black vertical line is $3\pi/2$ (left fig.) and BF (right fig.) and the dotted vertical red ones 3$\sigma$ interval. The horizontal pink line shows the upper bound on cosmological constraint $\sum_im_i=120$ meV. The orange band corresponds to the upper bound on $\langle m_{ee}\rangle$.
In the case of NH, $\delta$ is localized at [4.65-3$\pi/2$]$(\le3\pi/2)$ that is within the expected range of the analytical estimation in Eq.(\ref{eq:1dirac}).
But we obtain [4.705-4.72]$\in\delta$ in IH that implies we would need terms of next-to-next leading order of $s_{13}$ in Eq.(\ref{b1rel1}).
We predict $[0-0.015]\in\rho$, $[-0.10- 0]\in\sigma$, $[0.406-0.498]\in s^2_{23}$ in NH and
 $[-0.05-0]\in\rho$, $[0- 0.02]\in\sigma$, $[0.501-0.623]\in s^2_{23}$ in IH.
The sum of neutrino masses does not satisfy the bound on the current cosmological constant for both cases but this bound would easily be relaxed depending on experiments. The IH case is in favor of the BF on $s^2_{23}$. The neutrinoless double beta decay may be well-testable in the near future since the model results largely overlap its current bound.

\begin{figure}[tb]
\begin{center}
\includegraphics[width=70.0mm]{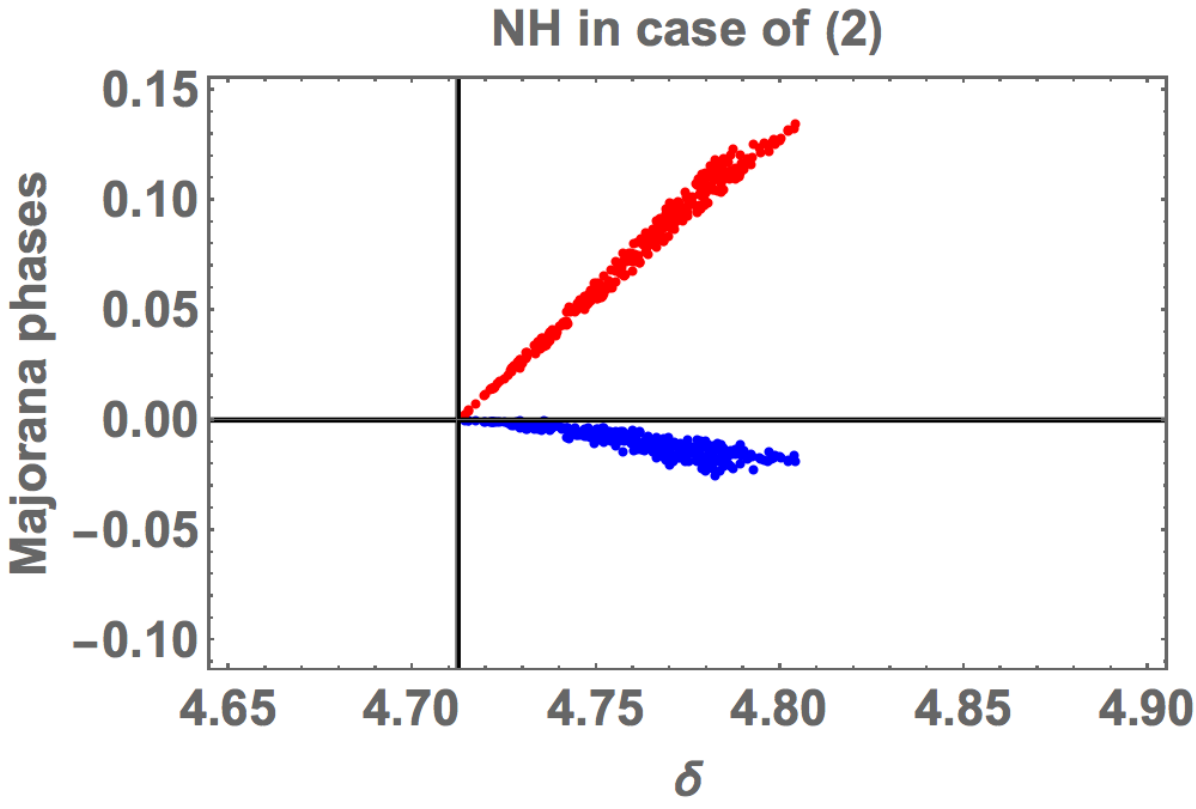} \quad
\includegraphics[width=70.0mm]{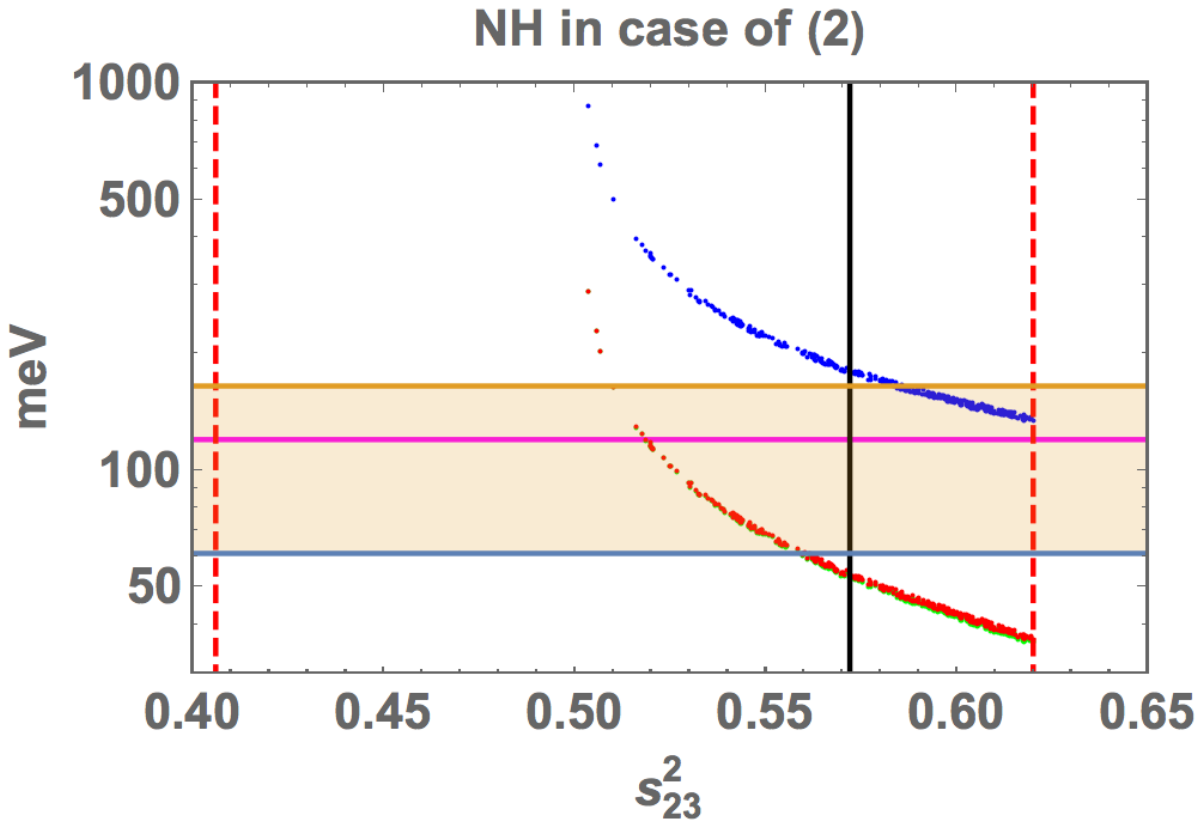}  \\
\includegraphics[width=70.0mm]{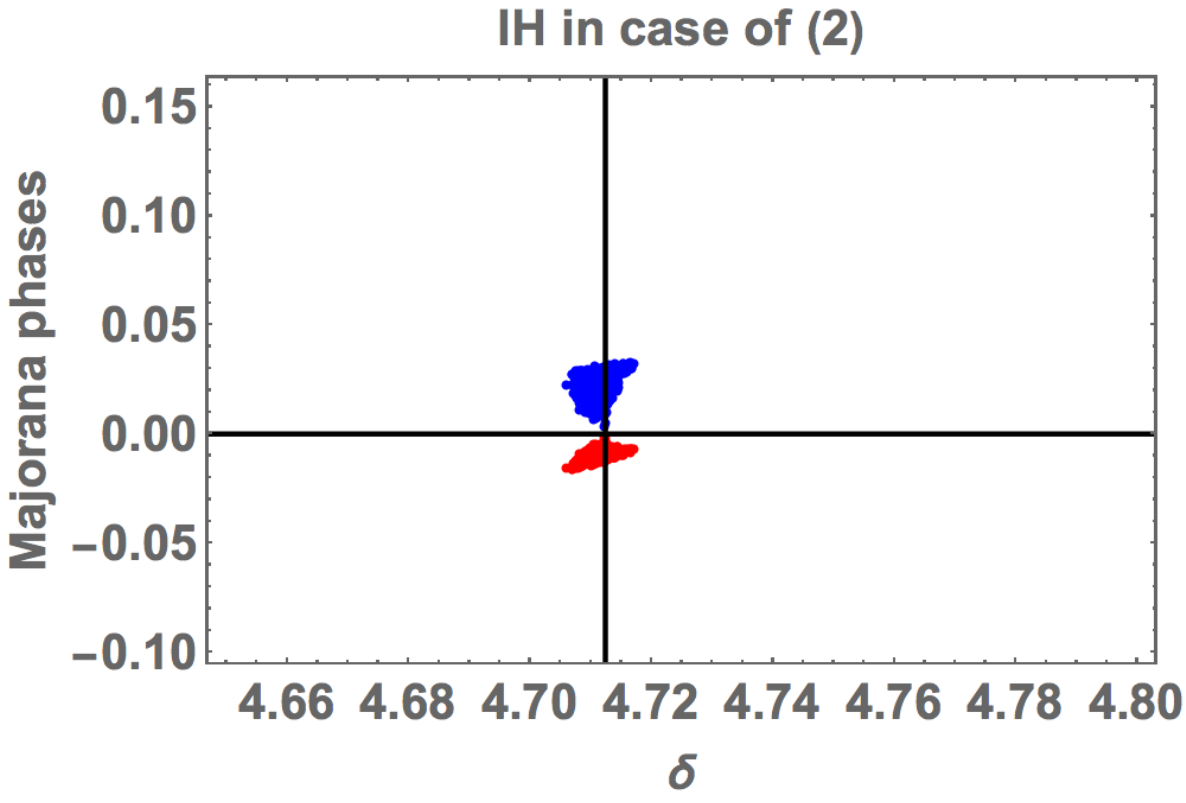} \quad
\includegraphics[width=70.0mm]{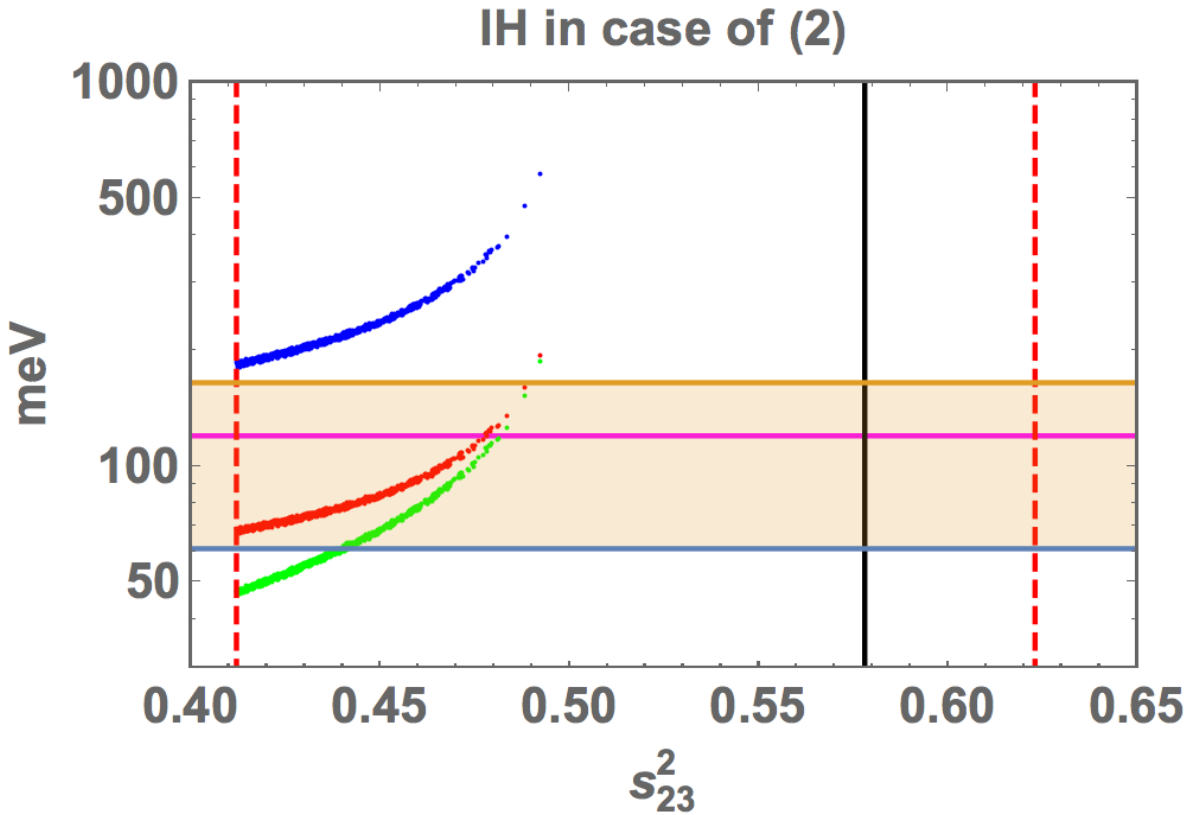} 
\caption{Numerical analyses of texture (2), where all the legends are the same as the ones of Fig.~\ref{fig:B1}.}
  \label{fig:B2}
\end{center}\end{figure}
%
\subsection{Texture (2)}
In Fig.~\ref{fig:B2}, we show predictions in the case of texture (2), where all the legends are the same as the case of Fig.~2.
In the case of NH, $\delta$ is localized at ($3\pi/2\le$)[3$\pi/2$-4.81] that is within the expected range of the analytical estimation in Eq.(\ref{eq:1dirac}).
But we obtain [4.705-4.72]$\in\delta$ in IH that implies we would need terms of next-to-next leading order of $s_{13}$ in Eq.(\ref{b1rel1}).
We predict $[-0.02-0]\in\rho$, $[0- 0.14]\in\sigma$, $[0.50-0.62]\in s^2_{23}$ in NH and
 $[-0.05-0]\in\rho$, $[0- 0.02]\in\sigma$, $[0.412-0.50]\in s^2_{23}$ in IH.
The sum of neutrino masses does not satisfy the bound on the current cosmological constraint for both cases, and NH case is in favor of the BF on $s^2_{23}$. The neutrinoless double beta decay may be well-testable in the near future since the model results largely overlap its current bound. Whole the nature of masses versus $s^2_{23}$ are also similar to the texture (1) when we shift $\theta_{23}$ to $\theta_{23}-\pi$ in Eq.~(\ref{eq:rel12}).

\subsection{Implication to phenomenology}

Here we briefly discuss phenomenology of the model focusing on doubly charged scalar boson $k^{\pm \pm}$.
The doubly charged scalar boson can be produced at collider experiments such as the LHC and future lepton colliders if its mass is $\mathcal{O}$(1) TeV scale.
For example, $k^{++}k^{--}$ pair can be produced via electroweak interaction at the LHC as $pp \to Z^*/\gamma^* \to k^{++}k^{--}$.
In fact the current upper limit of doubly charged scalar boson is around 900 GeV when it decays into charged leptons~\cite{ATLAS:2022pbd}.

The doubly charged scalar boson decays into the same sign dilepton via Yukawa interaction and the width is 
\begin{equation}
\Gamma(k^{\pm \pm} \to \ell^\pm \ell'^\pm) = \frac{|g_{\ell \ell'}|^2}{16 \pi (1 + \delta_{\ell \ell'})} m_{k^{\pm \pm}}.
\end{equation}
Thus the decay pattern of $k^{\pm \pm}$ is completely determined by the structure of coupling $g_{\ell \ell'}$.
In the model we can predict the decay branching ratio due to two-zero texture of $g_{\ell \ell'}$. 
In texture (1) $k^{\pm \pm}$ can only decay into $\{e^\pm e^\pm, e^\pm \mu^\pm, \mu^\pm \tau^\pm, \tau^\pm \tau^\pm \}$ modes. 
On the other hand, in texture (2), $k^{\pm \pm}$ can only decay into $\{e^\pm \mu^\pm, \mu^\pm \mu^\pm, e^\pm \tau^\pm, \tau^\pm \tau^\pm \}$ modes.
Therefore our model can be tested if the branching ratios are measured at collider experiments. 
Moreover Yukawa interactions can be tested by measuring electron scattering process at future electron colliders such as the ILC~\cite{Nomura:2017abh} and muon scattering process at muon colliders~\cite{Dev:2023nha} if the corresponding Yukawa couplings are not very small.
Detailed analysis of the collider phenomenology is beyond the scope of this paper and it will be discussed elsewhere.

\section{Summary and discussion}
\label{sec:V}
We have discussed verifiable two-zero textures in the framework of a three-loop induced neutrino mass model in which we have applied a modular $A_4$ symmetry and
found the mass structure is determined by assignments of the right-handed charged-lepton particles under the symmetry. 
We have shown three possible two-zero textures of the neutrino mass matrix. 
After performing their analytical and numerical analyses, we have discussed the possibility of detection for doubly-charged boson via colliders that decays into specific modes due to the flavor symmetries.

\section*{Acknowledgments}

The work was supported by the Fundamental Research Funds for the Central Universities (T.~N.), JSPS KAKENHI Grant Numbers JP20K14477 (H. Otsuka) and JP23H04512 (H. Otsuka).

\bibliography{ctma4.bib}
\end{document}